\documentclass[12pt]{article}

\begin{document}

\title{Gluon and quark distributions in large $N_{c}$ QCD:
theory vs. phenomenology}
\author{A.V. Efremov$^{a}$, K. Goeke$^{b}$ and P.V. Pobylitsa$^{c}$}
\date{}
\maketitle

\begin{center}
{\em \small
$^{a}$Joint Institute for Nuclear Research, Dubna, 141980, Russia
\\
$^{b}$Institut f\"{u}r Theoretische Physik II,
Ruhr-Universit\"{a}t Bochum,\\
D-44780 Bochum, Germany
\\
$^{c}$Petersburg Nuclear Physics Institute, Gatchina, St. Petersburg,
188350, Russia}
\end{center}

\begin{abstract}
We present the QCD analysis of the behavior of quark and gluon distributions
in the limit of a large number of colors $N_{c}$. We show that the results
agree qualitatively with the phenomenological data. In particular we draw
attention to the fact that the polarized gluon distribution is suppressed at
large $N_{c}$ compared to the unpolarized one.
\end{abstract}

The limit of large number of colors of QCD $N_{c}\rightarrow \infty $
and the $1/N_{c}$ expansion \cite{Hooft-74} are rather useful theoretical tools
for analyzing the nonperturbative effects of QCD. Although in the real world
we have $N_{c}=3$ which cannot be considered as a too large parameter, in
many cases the large $N_{c}$ limit is helpful for understanding the dynamics
of strong interactions. In particular, it was found that at large $N_{c}$
baryons can be described by a mean field solution of some effective meson
action \cite{Witten-79}. Although the exact form of this action (depending
on bilocal meson fields) is not known, certain conclusions can be made on
the large $N_{c}$ behavior of various quantities.

In particular, in the limit of large number of QCD colors $N_{c}$ the
twist-two parton distributions in nucleon typically behave as
\begin{equation}
f(x)=N_{c}^{2}\phi (N_{c}x)\,.  \label{f-leading}
\end{equation}
Here the function $\phi (z)$ is stable in the large $N_{c}$ limit (it has
the order $O(N_{c}^{0})$ and its argument $z=N_{c}x$ is assumed to be fixed
in the large $N_{c}$ limit).

One way to derive the large $N_{c}$ behavior (\ref{f-leading}) is to check
that the moments of parton distributions corresponding to (\ref{f-leading})
\begin{equation}
M_{j}=\int\limits_{-1}^{1}dx\,x^{j-1}f(x)=N_{c}^{-j+2}\int%
\limits_{-N_{c}}^{N_{c}}dy\,y^{j-1}\phi (y)=O(N_{c}^{-j+2})
\label{M-j-QCD-1}
\end{equation}
behave exactly as the matrix elements of the twist two operators $O_{\mu
_{1}\ldots \mu _{j}}$%
\begin{equation}
M_{j}=n^{\mu _{1}}\ldots n^{\mu _{j}}\langle N,p|O_{\mu _{1}\ldots \mu
_{j}}|N,p\rangle =O(N_{c}^{-j+2})\,.  \label{M-j-QCD}
\end{equation}
Here $n$ is a light-cone vector $n_{\mu }n^{\mu }=0$ normalized by condition
\begin{equation}
(np)=1\,.  \label{n-normalization}
\end{equation}
Taking nucleon at rest $p=(m_{N},0,0,0)$ and keeping in mind that the mass
of the nucleon at large $N_{c}$ behaves as
\begin{equation}
m_{N}=O(N_{c})
\end{equation}
we see from (\ref{n-normalization}) that the nonvanishing components of $n$
behave as $n=O(N_{c}^{-1})$ so that $j$ factors $n^{\mu _{1}}n^{\mu
_{2}}\ldots n^{\mu _{j}}$ contribute explicitly $O(N_{c}^{-j})$ to the
behavior of (\ref{M-j-QCD}) and in order to derive (\ref{M-j-QCD}) we must
check that
\begin{equation}
\langle N,p|O_{\mu _{1}\ldots \mu _{j}}|N,p\rangle =O(N_{c}^{2})\,.
\label{N-O-N-order}
\end{equation}
This can be done in a straightforward way using the rules of large $N_{c}$
counting \cite{Hooft-74}, \cite{Witten-79}. For example, in the case of the
quark operators $O_{\mu _{1}\ldots \mu _{j}}$ one order of $N_{c}$ comes
from the summation over color indices in operator $O_{\mu _{1}\ldots \mu
_{j}}$ and another order of $N_{c}$ is due to the relativistic normalization
of the nucleon state
\begin{equation}
\langle N,p^{\prime }|N,p\rangle =2p^{0}\delta ^{(3)}({\bf p}-{\bf p}
^{\prime })\,.
\end{equation}
Here for a nucleon at rest we have $p^{0}=m_{N}=O(N_{c})$ which gives the
second order of $N_{c}$ contributing to (\ref{N-O-N-order}). Thus (\ref
{M-j-QCD}) is checked.

Alternatively one can derive the large $N_{c}$ behavior (\ref{f-leading}) of
parton distributions directly following the logic of paper \cite{DPPPW-96}
(the derivation of \cite{DPPPW-96} was formulated in terms of the chiral
quark soliton model but its generalization to the ``true'' effective action
of large $N_{c}$ QCD formulated in terms of bilocal meson fields is
straightforward).

As an illustration of the large $N_{c}$
behavior (\ref{f-leading}) one can take
the baryon number sum rule
\begin{equation}
\int\limits_{0}^{1}\left[ u(x)+d(x)-\bar{u}(x)-\bar{d}(x)\right] dx=N_{c}
\label{b-number-sr}
\end{equation}
which obviously agrees with the large $N_{c}$ structure (\ref{f-leading}) of
quark distributions and with the behavior of moments (\ref{M-j-QCD-1}).

The physics standing behind the large $N_{c}$ structure (\ref{f-leading})
becomes rather transparent in the nonrelativistic quark model \cite
{Manohar-84}. Naively the nucleon consists of $N_{c}$ quarks and it is
natural to expect that its momentum (in the infinite momentum frame) is more
or less proportionally shared by all $N_{c}$ quarks. This explains why we
keep $N_{c}x$ fixed in the limit of large $N_{c}$ in (\ref{f-leading}). The
prefactor of $N_{c}^{2}$ in (\ref{f-leading}) is fixed e.g. by the
normalization condition (\ref{b-number-sr}).

As for the spin structures in this model, the proton consists of
$(N_{c}+1)/2$ $u$-quarks and $(N_{c}-1)/2$ $d$-quarks. The colored part of
the wave function is antisymmetric. Hence the spin-isospin wave function
must be symmetric (we assume that all quarks have the same orbital momentum
$l=0$). Hence the spin wave function of all $u$ quarks is symmetric and the
$(N_{c}+1)/2$ quarks have the maximal total spin $S_{u}$. The same is valid
for the spin of $(N_{c}-1)/2$ $d$-quarks, i.e.
\begin{equation}
S_{u}=\frac{1}{4}(N_{c}+1)\,,\quad S_{d}=\frac{1}{4}(N_{c}-1)\,.
\end{equation}
Interpreting the spin carried by $u$ and $d$ quarks as the moments of
polarized quark distributions we see that
\begin{equation}
\int\limits_{0}^{1}dx\Delta u(x)\sim \int\limits_{0}^{1}dx\Delta
d(x)dx=O(N_{c})
\end{equation}
in agreement with the general structure of the parton distributions (\ref
{f-leading}).

In order to obtain the proton state with spin $S=1/2$ and isospin $T=1/2$, $%
S_{u}$ and $S_{d}$ must be opposite to each other so that $O(N_{c})$ part of
the spin is cancelled and we are left with
\begin{equation}
\int\limits_{0}^{1}\left[ \Delta u(x)+\Delta d(x)\right] dx=O(N_{c}^{0})\,.
\end{equation}
This means that in the large $N_{c}$ limit function $\Delta u+\Delta d$ is
suppressed compared to (\ref{f-leading}) and has the form
\begin{equation}
\Delta u(x)+\Delta d(x)=N_{c}\phi ^{\prime }(N_{c}x)\,.
\label{singlet-polarized-suppressed}
\end{equation}
The above derivation of this suppression was rather heuristic: we checked
only the first moment and even this moment was estimated relying on the
naive quark model. Howefer, a straightforward QCD analysis reproduces this
result. One way to see it is to note that behavior (\ref
{singlet-polarized-suppressed}) was derived in \cite{DPPPW-96} in the
context of chiral models but can be trivially generalized to the ``true
meson effective action'' of large $N_{c}$ QCD. Alternatively one can check
the behavior of the moments in the spirit of \cite{Dashen-94}. Note that in
both approaches the suppression (\ref{singlet-polarized-suppressed}) is
essentially due to the so called spin-flavor symmetry
\cite{Witten-83,Dashen-94} of the mean field solution for the baryon
(the mean field approximation is justified at large $N_{c}$).

Actually not only $\Delta u+\Delta d$ but also some other flavor and
polarization {\em combinations} of parton distributions have this
suppressed large $N_{c}$ behavior:

\begin{equation}
f_{\rm subleading}(x)=N_{c}\phi ^{\prime }(N_{c}x)\,.  \label{subleading}
\end{equation}
For example, the isovector unpolarized distribution $u-d$ is also of this
type, which can be seen from the isospin sum rule
\begin{equation}
\int\limits_{0}^{1}\left\{ \left[ u(x)-d(x)\right] -[\bar{u}(x)-\bar{d}
(x)]\right\} dx=1\,.
\end{equation}

In the table we collect the relevant combinations of twist-two parton
distributions. The main result of the large $N_{c}$ counting is that the
isoscalar unpolarized and isovector polarized distributions are dominant in
the $1/N_{c}$ limit whereas the isovector unpolarized and isoscalar
polarized are suppressed (within the chiral quark soliton model this was
established in \cite{DPPPW-96,PP-96}, the generalization to the large $N_{c}$
QCD is straightforward).

\vspace{0.5cm}

\begin{center}
\begin{tabular}{|l|l|l|}
\hline
parton distributions &
\begin{tabular}{l}
leading \\
$N_{c}^{2}\phi (xN_{c})$%
\end{tabular}
&
\begin{tabular}{l}
suppressed \\
$N_{c}\phi ^{\prime }(xN_{c})$%
\end{tabular}
\\ \hline
\hspace{0.4em}unpolarized & $u+d$ & $u-d$ \\ \hline
\begin{tabular}{l}
longitudinally \\
polarized
\end{tabular}
& $\Delta u-\Delta d$ & $\Delta u+\Delta d$ \\ \hline
\hspace{0.4em}transversity & $\delta u-\delta d$ & $\delta u+\delta d$ \\
\hline
\hspace{0.4em}gluon & $G$ & $\Delta G$ \\ \hline
\end{tabular}
\end{center}

\vspace{0.5cm}

Note that for the antiquark distributions the large $N_{c}$ counting is the
same as for the quark ones (at this point instead of the nonrelativistic
quark model it is better to rely on the arguments of \cite{DPPPW-96} or on
the direct QCD large $N_{c}$ counting).

As for the gluon distribution, its large $N_{c}$ behavior can be estimated
simply by generating the gluons through evolution from quarks. The one-loop
DGLAP kernel $\alpha _{s}P_{qg}$ corresponding to this radiation is of order
\begin{equation}
\alpha _{s}P_{qg}\sim \alpha _{s}C_{F}=\alpha _{s}\frac{N_{c}^{2}-1}{2N_{c}}%
\sim \alpha _{s}N_{c}=O(N_{c}^{0})
\end{equation}
since $\alpha _{s}=O(N_{c}^{-1})$. This means that the gluon distributions
inherit the large $N_{c}$ behavior of isoscalar quark distributions (\ref
{f-leading}). It should be stressed that here we keep in mind the domain $%
x\sim 1/N_{c}$ of ``small but not too small $x$''. The region corresponding
to the physics of really small $x$ is beyond our consideration.

Now let us turn to the phenomenological consequences of the large $N_{c}$
counting. For the parton distributions whose shape is determined with
a certain reliability ($u,d,\Delta u,\Delta d$) the large $N_{c}$ behavior
\begin{equation}
u-d\sim \frac{1}{N_{c}}(u+d)\,,  \label{u-d-N-c}
\end{equation}
\begin{equation}
\Delta u+\Delta d\sim \frac{1}{N_{c}}(\Delta u-\Delta d)
\label{Delta-u-min-Delta-d-N-c}
\end{equation}
really seems to be in agreement with the experimental data.

Indeed, integrating both sides of (\ref{u-d-N-c}) (weighted with $x$) and (%
\ref{Delta-u-min-Delta-d-N-c}) over $x$ we arrive with the recent global fit
of unpolarized data \cite{MRST} at $Q^{2}=2$GeV$^{2}$ at
\begin{equation}
\frac{\int_{0}^{1}dx\,x\left[ (u-d)+(\bar{u}-\bar{d})\right] }{
\int_{0}^{1}dx\,x\left[ (u+d)+(\bar{u}+\bar{d})\right] }
\approx \frac{1}{3.5}
\end{equation}
and for polarized data at
\begin{equation}
\frac{\int_{0}^{1}dx\left[ (\Delta u+\Delta d)+(\Delta \bar{u}+\Delta \bar{d}%
)\right] }{\int_{0}^{1}dx\left[ (\Delta u-\Delta d)+(\Delta \bar{u}-\Delta
\bar{d})\right] }=\frac{2\Delta \Sigma +a_{8}}{3g_{A}}\approx \frac{1}{2.8}
\end{equation}
Here we used the values $g_{A}=1.26$, $a_{8}=0.58$, $\Delta \Sigma =0.39$
\cite{Leader-99}. The results agree with the $O(1/N_{c})$ suppression (\ref
{u-d-N-c}), (\ref{Delta-u-min-Delta-d-N-c}).

An interesting result of the above analysis is that $\Delta G$ is $1/N_{c}$
suppressed compared to $G$. A naive (but still justified for $1/N_{c}$
counting) way to understand this is again to use the radiative mechanism of
the gluon generation: since the polarized $\Delta u+\Delta d$ distribution
is $1/N_{c}$ suppressed in the sense of (\ref{subleading}) we expect that $%
\Delta G$ has the same large $N_{c}$ behavior. A more scientific way to
derive the large $N_{c}$ suppression of $\Delta G$ would be to make use of
the fact that in the large $N_{c}$ limit the nucleon can be effectively
described by a sort of Hartree (mean field) approach in terms of bilocal
meson fields (including the composite fields made of gluons). The mean field
appearing in this Hartree approach is believed to have mixed
spin-flavor symmetry \cite{Witten-83,Dashen-94}.
In the gluon sector this symmetry of the mean field solution
simply reduces to the pure rotational symmetry which leads to the
suppression of the polarized gluon distribution.

Thus
\begin{equation}
\frac{\Delta G}{G}\sim \frac{1}{N_{c}}\,.
\end{equation}
This is consistent with the first direct experimental estimation \cite
{HERMES-2000} of this ratio in the interval $0.06<x<0.28$%
\begin{equation}
\frac{\int_{0.06}^{0.28}dx\Delta G}{\int_{0.06}^{0.28}dxG}=0.41\pm 0.18
\end{equation}

In the case of antiquark polarized distributions we have
\begin{equation}
\Delta \bar{u}+\Delta \bar{d}\sim \frac{1}{N_{c}}(\Delta \bar{u}-\Delta \bar{%
d})\,.
\end{equation}
This leads us to the conclusion that in the large $N_{c}$ limit
\begin{equation}
\Delta \bar{u}\sim -\Delta \bar{d}\,.
\end{equation}
This is supported by calculations in the chiral quark soliton
model \cite{DPPPW-96,Dressler-99} where the large flavor asymmetry $\Delta
\bar{u}-\Delta \bar{d}$ was
obtained (for the numerical parametrization see
\cite{GPPU-00}). Also note that this result agrees
with the instanton model prediction \cite{DK},
with the phenomenological
approach of \cite{Glueck-Reya-2000} based on the ``Pauli blocking
principle'' and with the statistical model of \cite{Bhalerao-00}.
Our opinion is that at the current level of the uncertainties the
large $N_{c}$ behavior $\Delta \bar{u}\approx -\Delta \bar{d}$ seems more
reliable than the {\em ad hoc}
assumption $\Delta \bar{u}=\Delta \bar{d}$ which is
often used in the analysis of phenomenological data.
An indication of a nonzero value for
$\Delta \bar{u} -\Delta \bar{d}$  was also observed in
\cite{MY}.

Concerning the transversity distribution, the large $N_{c}$ QCD predicts $%
\delta u=-\delta d$ up to $1/N_{c}$ corrections (see the table). The
different signs of $\delta u$ and $\delta d$ manifest themselves in the
opposite signs of $\pi ^{+}$ and $\pi ^{-}$ azimuthal spin asymmetries
in semi-inclusive DIS \cite{EGPU-00} and in high $p_{T}$ inclusive pion
production \cite{Anselmino-99} which are observed experimentally \cite
{HERMES-99,Adams-91}.

Certainly, the parameter $1/N_{c}=1/3$ is not too small and looking at a
specific quantity one cannot unambiguously conclude whether we really
observe the $1/N_{c}$ suppression or not. However, the above examples (see
also \cite{PP-Soffer} for the analysis of the large $N_{c}$ bounds on
transversity distributions) give a certain evidence that the experimental
data for parton distributions are in a rather reasonable agreement with the
large $N_{c}$ counting of QCD. As a particular result of this investigation
one can assert that the polarized gluon distribution is suppressed by $%
1/N_{c}$ compared to the unpolarized one.

{\bf Acknowledgment.} We are grateful to M. Gl\"{u}ck, V.Yu. Petrov, M.V.
Polyakov, E. Reya, A. Shuvaev and O. Teryaev for discussions.
This work was partially supported by RFBR Grant 00-02-16696, by DFG and by
BMBW. A.E. and P.P. are thankful to the Institute for Theoretical Physics of
Ruhr University Bochum for warm hospitality.

\end{document}